\documentclass[aps,prl,twocolumn,groupedaddress,superscriptaddress,showpacs,msmath,amssymb]{revtex4}

\usepackage{graphicx}
\usepackage{dcolumn}
\usepackage{bm}
\usepackage[usenames]{color}
\usepackage[normalem]{ulem}

\begin{document}

\title{
Experimental verification of PbBi$_{2}$Te$_{4}$ as a 3D topological insulator
}

\author{K. Kuroda}
\affiliation{%
Graduate School of Science, Hiroshima University, 1-3-1 Kagamiyama, Higashi-Hiroshima 739-8526, Japan\\
}

\author{H. Miyahara}
\affiliation{%
Graduate School of Science, Hiroshima University, 1-3-1 Kagamiyama, Higashi-Hiroshima 739-8526, Japan\\
}

\author{M. Ye}
\affiliation{%
Graduate School of Science, Hiroshima University, 1-3-1 Kagamiyama, Higashi-Hiroshima 739-8526, Japan\\
}

\author{S. V. Eremeev}
\affiliation{%
Institute of Strength Physics and Materials Science, 634021, Tomsk, Russia\\
}
\affiliation{%
Tomsk State University, 634050, Tomsk, Russia\\
}

\author{Yu. M. Koroteev}
\affiliation{%
Institute of Strength Physics and Materials Science, 634021, Tomsk, Russia\\
}
\affiliation{%
Tomsk State University, 634050, Tomsk, Russia\\
}

\author{E. E. Krasovskii}
\affiliation{%
Departamento de F\'{\i}sica de Materiales UPV/EHU
and Centro de F\'{\i}sica de Materiales CFM
and Centro Mixto CSIC-UPV/EHU,
20080 San Sebasti\'an/Donostia, Basque Country, Spain
\\
}
\affiliation{%
Donostia International Physics Center (DIPC), 20018 San
Sebasti\'an/Donostia, Basque Country, Spain\\
}
\affiliation{%
IKERBASQUE, Basque Foundation for Science, 48011 Bilbao, Spain\\
}

\author{E. V. Chulkov}
\affiliation{%
Departamento de F\'{\i}sica de Materiales UPV/EHU
and Centro de F\'{\i}sica de Materiales CFM
and Centro Mixto CSIC-UPV/EHU,
20080 San Sebasti\'an/Donostia, Basque Country, Spain
\\
}
\affiliation{%
Donostia International Physics Center (DIPC),
             20018 San Sebasti\'an/Donostia, Basque Country,
             Spain\\
}

\author{S. Hiramoto}
\affiliation{%
Graduate School of Science, Hiroshima University, 1-3-1 Kagamiyama, Higashi-Hiroshima 739-8526, Japan\\
}

\author{C. Moriyoshi}
\affiliation{%
Graduate School of Science, Hiroshima University, 1-3-1 Kagamiyama, Higashi-Hiroshima 739-8526, Japan\\
}

\author{Y. Kuroiwa}
\affiliation{%
Graduate School of Science, Hiroshima University, 1-3-1 Kagamiyama, Higashi-Hiroshima 739-8526, Japan\\
}

\author{K. Miyamoto}
\affiliation{
Hiroshima Synchrotron Radiation Center, Hiroshima University, 2-313 Kagamiyama, Higashi-Hiroshima 739-0046, Japan\\
}

\author{T. Okuda}
\affiliation{
Hiroshima Synchrotron Radiation Center, Hiroshima University, 2-313 Kagamiyama, Higashi-Hiroshima 739-0046, Japan\\
}

\author{M. Arita}
\affiliation{
Hiroshima Synchrotron Radiation Center, Hiroshima University, 2-313 Kagamiyama, Higashi-Hiroshima 739-0046, Japan\\
}

\author{K. Shimada}
\affiliation{
Hiroshima Synchrotron Radiation Center, Hiroshima University, 2-313 Kagamiyama, Higashi-Hiroshima 739-0046, Japan\\
}

\author{H. Namatame}
\affiliation{
Hiroshima Synchrotron Radiation Center, Hiroshima University, 2-313 Kagamiyama, Higashi-Hiroshima 739-0046, Japan\\
}

\author{M. Taniguchi}
\affiliation{%
Graduate School of Science, Hiroshima University, 1-3-1 Kagamiyama, Higashi-Hiroshima 739-8526, Japan\\
}
\affiliation{
Hiroshima Synchrotron Radiation Center, Hiroshima University, 2-313 Kagamiyama, Higashi-Hiroshima 739-0046, Japan\\
}

\author{Y. Ueda}
\affiliation{
Kure National College of Technology, Agaminami 2-2-11, Kure 737-8506, Japan\\
}

\author{A. Kimura}
\email{akiok@hiroshima-u.ac.jp}
\affiliation{%
Graduate School of Science, Hiroshima University, 1-3-1 Kagamiyama, Higashi-Hiroshima 739-8526, Japan\\
}

\date{\today}

\begin{abstract}
The first experimental evidence is presented of the topological insulator
state in PbBi$_{2}$Te$_{4}$. A single surface Dirac cone is observed by
angle-resolved photoemission spectroscopy (ARPES) with synchrotron radiation. 
Topological invariants $\mathbb{Z}_2$ are calculated from the {\it ab initio}
band structure to be 1; (111). The observed two-dimensional iso-energy contours 
in the bulk energy gap are found to be the largest among the known three-dimensional 
topological insulators. This opens a pathway to achieving a sufficiently large 
spin current density in future spintronic devices.
\end{abstract}


\maketitle

\begin{figure*} [t]
\includegraphics{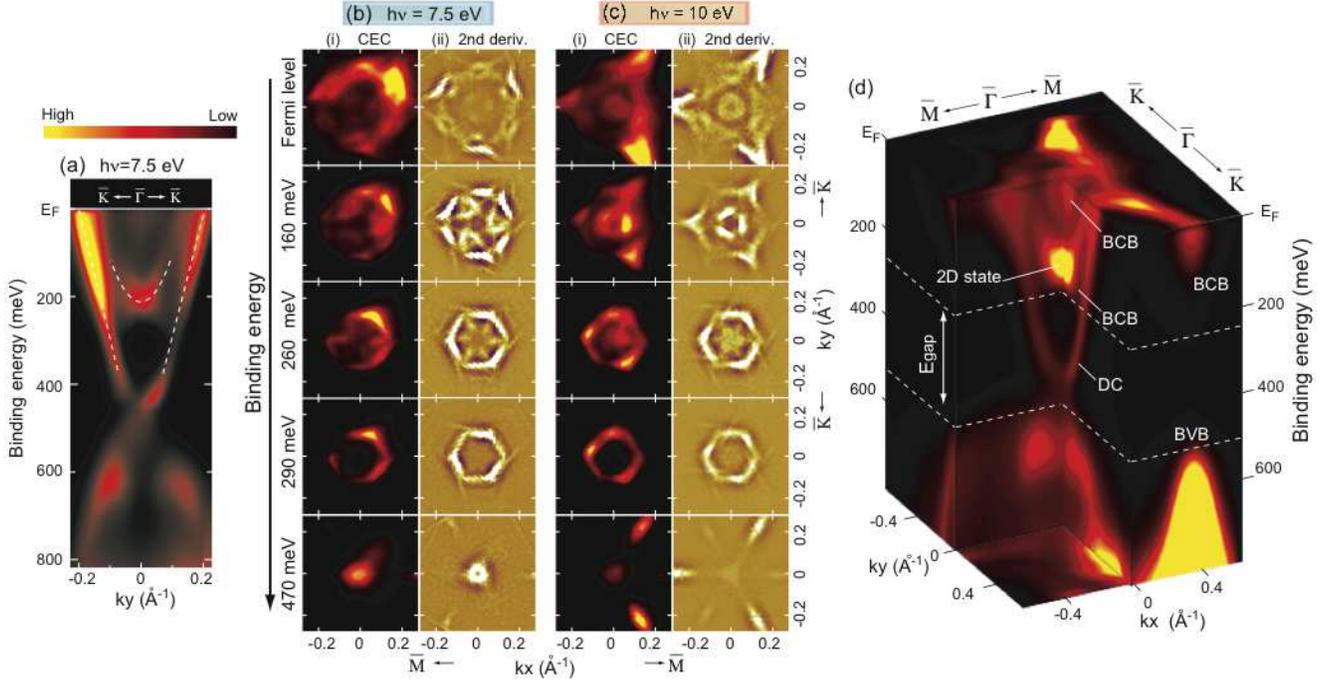}
\caption{(Color online) 
(a) ARPES energy dispersion curves along the $\bar{\Gamma}\bar{\rm K}$ line 
acquired at $h\nu=7.5$~eV. Constant energy contour maps (i) and their 
second derivatives (ii) obtained from the 
ARPES measurement at $h\nu=7.5$~eV (b) and 10~eV (c) in the ${\mathbf k}_\parallel$ 
range 
$-0.3$~\AA$^{-1}\le k_{x}, k_{y}\le +0.3$~\AA$^{-1}$
from $E_{\rm B}=470$~meV (Dirac point) to 0 (Fermi level). The three-fold 
symmetrization procedure is applied only for the second derivative images.  
(c) Three dimensional map for $h\nu=10$~eV.  
}
\end{figure*}

Topological insulators (TIs) have recently emerged as a new state of quantum
matter, which are distinguished from conventional insulators by having
a surface state with a massless Dirac dispersion in the bulk energy gap. 
The spin orientation of the topological surface state is locked with its 
crystal momentum, resulting in a helical spin texture. This new state 
can be classified by so-called $\mathbb{Z}_2$ topological invariants~\cite{FKM_07,FK_07}. 
Generally, the massless Dirac cone can be created at an interface of two 
materials: a topological and an ordinary insulator. The unique properties 
of topological surface electrons provide a fertile ground to realize new 
electronic phenomena, such as a magnetic monopole arising from the topological 
magneto-electric effect and Majorana fermions at the interface with a superconductor
~\cite{Hasan&Kane_RMP,Qi&Zhang_RMP}. Owing to time-reversal symmetry, topological 
surface states are protected from backscattering in the presence of a weak 
perturbation, which is required for the realization of dissipationless spin 
transport without external magnetic fields in novel quantum devices~\cite{Xue_11,Xiu_11}.

A number of materials that hold spin-polarized surface Dirac cones
have been intensively studied, such as
Bi$_{1-x}$Sb$_{x}$~\cite{Hsieh_Science_09, Nishide_PRB_10},
Bi$_{2}$Te$_{3}$~\cite{Chen_Science_09,Hsieh_PRL2009},
Bi$_{2}$Se$_{3}$~\cite{Xia_NatPhys_09,
Kuroda_PRL_10_Bi2Se3,Bi2Se3_eph,Kim_PRL_11}, 
and thallium-based compounds
\cite{eremeev1,Eremeev_Tl_PRB2011,Kuroda_PRL_10_TlBiSe2,Yan_EPL_10,Lin_PRL_10,Sato_PRL_10,Chen_PRL_10}.
Among them, Bi$_{2}$Se$_{3}$ has been regarded as one of the most
promising candidates for potential applications in ultra-low power
consumption quantum devices that can work stably at room temperature
due to a sufficiently large bulk energy gap.
Although significant efforts have been made towards spintronic
applications, 
the surface contribution to conduction was hardly observed even at low
bulk carrier density~\cite{Analytis_PRB_10, Eto_PRB_10,Butch_PRB_10}. 
This motivates the search of new topological insulators with higher 
density of spin-polarized Dirac fermions.

Recently, some of the Pb-based ternary chalcogenides have been proposed as 
3D topological insulators~\cite{Eremeev_JETP_10,Jin_PRB_11,Menshchikova_JETP_11}.  
The present study is focused on PbBi$_{2}$Te$_{4}$, which has the simplest crystal 
structure among them. The crystal is composed of seven-layer blocks with the atomic 
layer sequence Te-Bi-Te-Pb-Te-Bi-Te [see Fig.~S1(a) of supplementary material (SM)]~\cite{Shelimova_PRB_11}.  
The theoretical analysis of Ref.~\cite{Menshchikova_JETP_11} indicates that this 
compound is a three-dimensional (3D) topological insulator. In this Letter, we 
report the first experimental evidence of the topological surface state in 
PbBi$_{2}$Te$_{4}$ by angle-resolved photoemission spectroscopy (ARPES).  
We establish that the size of the two-dimensional iso-energy contour, which 
is proportional to the surface carrier density, is the largest among the known 
3D topological insulators. Thus, PbBi$_{2}$Te$_{4}$ can be counted as one of 
the most promising candidates for realizing a large spin current density in 
future spintronic devices.

A single crystalline sample of PbBi$_{2}$Te$_{4}$ was grown by the
standard procedure using Bridgman method [see SM
for more details]. For ARPES measurement the samples 
were {\it in situ} cleaved along the basal plane. Photoemission experiment 
was performed with synchrotron radiation at the linear undulator beamline 
(BL1) and the helical undulator beamline (BL9A) of Hiroshima Synchrotron 
Radiation Center (HiSOR). The ARPES spectra were acquired with a hemispherical 
photoelectron analyzer (VG-SCIENTA R4000) at 17~K. The overall energy and 
angular resolutions were set to 10--20~meV and 0.3$^{\circ}$, respectively.

Figure 1(a) shows the ARPES energy dispersion curve along the
$\bar{\Gamma}\bar{\rm K}$ line of the surface Brillouin zone [see SM, Fig.~S1(b)] 
measured at $h\nu=7.5$~eV. The parabolic band with the energy minimum at the 
binding energy of $E_{\rm B}=200$~meV is seen to exhibit a rather strong 
photoemission intensity near the $\bar{\Gamma}$ point. The position of the 
energy minimum does not change with the photon energy [see SM, Figs.~S2(a) and S2(b)], 
which confirms the two-dimensional nature of this electronic state. More importantly, 
a linearly dispersing feature, i.e. the Dirac cone with the crossing point at 
$E_{\rm B}=470$~meV is observed.

Figures 1(b) and 1(c) summarize the constant energy contours (i) and their 
second derivatives (ii) in the ${\mathbf k}_\parallel$ 
range 
$-0.3$~\AA$^{-1}\le k_{x}, k_{y}\le +0.3$~\AA$^{-1}$ from $E_{\rm B}=470$~meV (Dirac point) to 0 (Fermi level)
at $h\nu=7.5$ and 10~eV, respectively. With $h\nu=10$~eV 
we find at the Dirac point energy six ellipses oriented along $\bar{\Gamma}\bar{\rm M}$ 
in  addition to the point-like feature at the $\bar{\Gamma}$ point for the Dirac
state. On the other hand, the elliptical contours are much weaker at $h\nu=7.5$~eV, 
signifying a strong matrix elements effect. In going away from the Dirac point, the 
sole hexagonally shaped contour is observed down to $E_{\rm B}=290$~meV and at
smaller binding energies another state becomes enclosed inside the Dirac cone  
(starting with $E_{\rm B}=260$~meV). It is interpreted as the onset of the bulk
conduction band (BCB) because the shape of the inner contours considerably depends 
on the photon energy, see Figs.~1(b) and 1(c). Below $E_{\rm B}=160$~meV the Dirac 
cone further deforms and practically merges into the bulk conduction band.
In the map measured with $h\nu=10$~eV the hexagonal Fermi surface of the 
Dirac cone encloses two large and one small triangular shaped surfaces centered
at $\bar{\Gamma}$. At $h\nu=7.5$~eV the shape of the inner Fermi surfaces 
strongly changes and becomes very complicated, which is consistent with a bulk state [see Fig.~S4 of SM].
The size of the energy gap is estimated as 230~meV as shown 
in the 3D map for $h\nu=10$~eV [Fig.~1(d)]. Such a large energy gap is beneficial for a high
stability of the spin current conductance at room temperature.
Figure 1(d) illustrates a high anisotropy of the bulk valence band: 
the valence band maximum in the $\bar{\rm M}\bar{\Gamma}\bar{\rm M}$ line is 
at $E_{B}=490$~meV ($k_{||}=\pm 0.3$~\AA$^{-1}$), while in the 
$\bar{\rm K}\bar{\Gamma}\bar{\rm K}$ line it is deeper in energy.

Next we discuss the surface Dirac cone of PbBi$_{2}$Te$_{4}$ in more detail 
by comparing it with other 3D TIs. The Dirac cone dispersion along
$\bar{\Gamma}\bar{\rm M}$ and $\bar{\Gamma}\bar{\rm K}$ is shown in Fig.~2(a) 
for PbBi$_{2}$Te$_{4}$ and for the well studied TIs Bi$_{2}$Se$_{3}$ and 
TlBiSe$_{2}$.
Close to $E_{\rm F}$ the Dirac cone energy dispersion in PbBi$_{2}$Te$_{4}$ 
is as steep as in the other materials, but it is apparently less steep near 
the Dirac point. In PbBi$_{2}$Te$_{4}$, the group velocity at $E_{\rm F}$ is 
estimated as $3.9\times 10^{5}$~m/s, while near the Dirac point, it is much 
lower ($1.4\times 10^{5}$~m/s) than in Bi$_{2}$Se$_{3}$ 
($2.9\times 10^{5}$~m/s)~\cite{Kuroda_PRL_10_Bi2Se3} and in TlBiSe$_{2}$ 
($3.9\times 10^{5}$~m/s)~\cite{Kuroda_PRL_10_TlBiSe2}.
The sizes of the iso-energy contours in the bulk energy gap from the 
Dirac point to 200~meV are much larger for PbBi$_{2}$Te$_{4}$ than for the
other two materials, as shown in Fig~.2(b). The estimated topological 
surface carrier density in PbBi$_{2}$Te$_{4}$ obtained from the area 
of constant energy contour $S(E)/4\pi^2$ is much larger than in the 
other two materials [Fig.~2(c)].

\begin{figure} [t]
\includegraphics{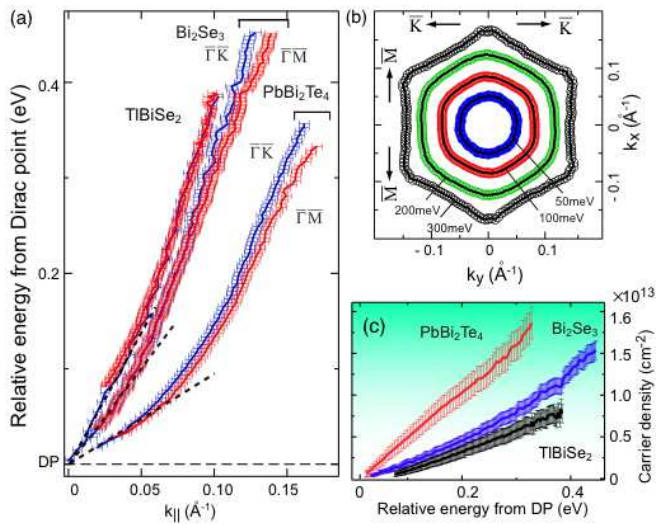}
 \caption{(Color online) (a) Experimental energy dispersion curves of
surface Dirac cones along $\bar{\Gamma}\bar{\rm M}$ and $\bar{\Gamma}\bar{\rm K}$ 
lines in PbBi$_{2}$Te$_{4}$ compared to Bi$_{2}$Se$_{3}$ and TlBiSe$_{2}$.  
The curves are obtained from the intensity maxima of momentum distribution 
curves. The energy is relative to the Dirac point. (b) Constant energy 
contours for PbBi$_{2}$Te$_{4}$. 
(c) Estimated carrier densities $S(E)/4\pi^2$ for three materials as a 
function of energy with respect to Dirac point. $S(E$) is the area of 
the constant energy contour at energy $E$. 
}
\end{figure}
\begin{figure} [t]
\includegraphics{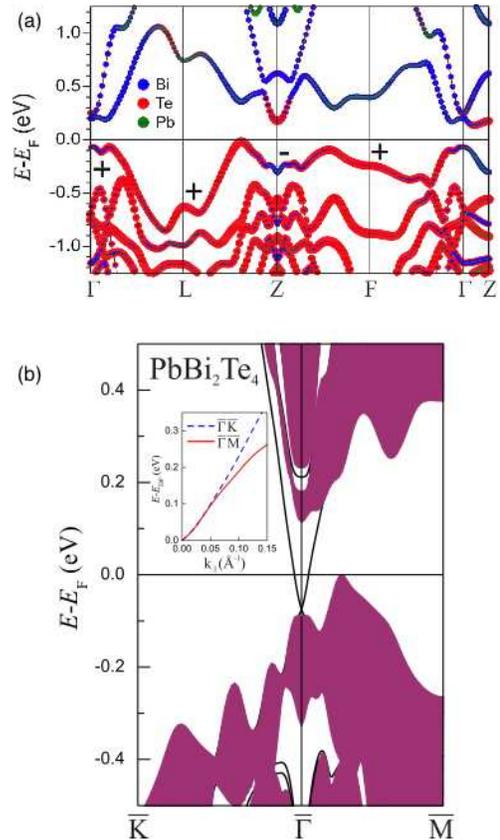}
 \caption{(Color online) (a) Bulk band structure of PbBi$_{2}$Te$_{4}$
calculated along high symmetry directions of the Brillouin zone;
colors show the weight of the states at Bi (blue), Te (red), and Pb
(green) atoms, respectively. Signs of $\delta_i=\pm 1$ at the TRIM
are also shown. (b) Calculated surface (solid line) and bulk continuum states (shaded)
of PbBi$_{2}$Te$_{4}$ along $\bar{\rm K}$-$\bar{\Gamma}$-$\bar{\rm M}$ line.
The inset shows the surface Dirac cones in the limited momentum space for 
$\bar{\Gamma}$-$\bar{\rm M}$ and $\bar{\Gamma}$-$\bar{\rm K}$ lines.
}
\end{figure}

The theoretical bulk and surface band structures of PbBi$_{2}$Te$_{4}$
are shown in Figs.~3(a) and 3(b), respectively. The calculations were
performed with the VASP code~\cite{VASP,PAW} with optimized internal 
lattice parameters.
In principle, three-dimensional materials with inversion symmetry
are classified with four $\mathbb{Z}_2$ topological invariants
$\nu_{0}; (\nu_{1}\nu_{2}\nu_{3})$, which can be determined by the parity 
$\xi_{m}(\Gamma_i)$ of occupied bands at eight time-reversal invariant 
momenta (TRIM)
$\Gamma_{i=(n_1,n_2,n_3)}=(n_1{\bf b}_1+n_2{\bf b}_2+n_3{\bf b}_3)/2$,
where ${\bf b}_1$, ${\bf b}_2$, ${\bf b}_3$ are primitive reciprocal 
lattice vectors, and $n_j=0$ or 1~\cite{FKM_07, FK_07}. 
The $\mathbb{Z}_2$ invariants are determined by the equations
$(-1)^{\nu_0}=\prod\limits_{i=1}^{8} \delta_{i}$ and
$(-1)^{\nu_k}=\prod\limits_{n_k=1;n_{j\neq k}=0,1} \delta_{i=(n_1n_2n_3)}$, 
where $\delta_{i}=\prod\limits_{m=1}^{N}\xi_{2m}(\Gamma_{i})$~\cite{FK_07}.
For rhombohedral lattice of PbBi$_{2}$Te$_{4}$ the TRIMs are $\Gamma$,
Z, and three equivalent L as well as F points [see SM, Fig.~S1(b)].
The previous study confirmed that this compound is a strong topological
insulator with the principal topological invariant $\nu_0=1$
\cite{Menshchikova_JETP_11}. 
Here we analyze the $(\nu_1\nu_2\nu_3)$ invariants.
Interestingly, the parity inversion of bulk bands occurs at the Z point for
PbBi$_{2}$Te$_{4}$ [Fig.~3(a)], which leads to $\mathbb{Z}_2$ invariants 
1; (111). This is in contrast to the case of binary chalcogenides Bi$_{2}$X$_{3}$ 
(X=Se, Te), where the parity inversion takes place at the $\Gamma$ point with 
$\mathbb{Z}_2$ invariant 1; (000)~\cite{Hasan&Kane_RMP}. Note that PbBi$_{2}$Te$_{4}$ 
is the first case among the experimentally established topological insulators 
with $\mathbb{Z}_2$ invariant 1; (111) possessing a single Dirac cone surface 
state. 
It is, thus, distinguished from the Bi$_{1-x}$Sb$_{x}$ alloy with the same
$\mathbb{Z}_2$ invariant~\cite{Teo_08} but with 5 or 3 pairs of surface states
crossing the Fermi energy~\cite{Hsieh_Science_09, Nishide_PRB_10}.
In Bi$_{1-x}$Sb$_{x}$, owing to nonzero invariants $(\nu_1\nu_2\nu_3)$, a 
one-dimensional (1D) topologically protected state can exist at the dislocation 
core~\cite{Ran_09}. In the case of the layered crystal the bulk dislocations can 
hardly exist, but other types of 1D TI states, such as edge states in thin films 
or 1D states at step edges are possible.

Finally, we compare the experimental ARPES results with the theoretical 
band structures, see Fig.~3(b). The calculated band structure well reproduces 
the bulk conduction band minimum located at $\bar{\Gamma}$ point at 200~meV 
above the Dirac point. Although the bulk valence band maximum is higher in 
theory than in experiment, the theory well reproduces its location in
${\bf k}$-space, that is, it appears around 1/3$\bar{\Gamma}\bar{\rm M}$ 
($k_{\parallel}\sim 0.3$~\AA$^{-1}$). Anisotropic features along different 
symmetry lines can be recognized above 100~meV relative to the Dirac point, 
as depicted in the inset. In addition, the parabolic surface state appears at
$\sim 280$~meV above the Dirac point in the conduction band gap, which is 
consistent with the present observation [see SM, Fig.S3].

Our conclusion led by the present experiment is twofold: (i) PbBi$_{2}$Te$_{4}$ 
is proved to be a three-dimensional topological insulator with the energy gap 
of 230~meV and a single Dirac cone at the $\bar{\Gamma}$ point. (ii) The size 
of the Fermi surface contour in the bulk energy gap is significantly larger  
than in the other presently known 3D topological insulators, whereby the 
highest carrier density of the known topological surface states is achieved.
These novel findings pave a way for the efficient control of the group
velocity with sufficiently large spin current density by tuning the
chemical potential in the bulk energy gap.

We thank Shuichi Murakami for valuable comments.
We also thank J.~Jiang, H.~Hayashi, T.~Habuchi and H.~Iwasawa
for their technical support in the ARPES measurement at
Hiroshima Synchrotron Radiation Center (HSRC).
This work was financially supported by
KAKENHI (Grant No. 20340092, 23340105), Grant-in-Aid for Scientific
Research (B) of JSPS.
We also acknowledge partial support by the Department of Education
of the Basque Country Government, the University of the Basque Country
(project GV-UPV/EHU, grant IT-366-07), Ministerio de Ciencia e Inovaci{\'o}n
(grant FIS2010-19609-C02-00).
Calculations were performed on SKIF-Cyberia
(Tomsk State University) and Arina (UPV/EHU) supercomputers.
The ARPES measurement was performed with the approval of the
Proposal Assessing Committee of HSRC (Proposal No.11-A-3, 11-A-4).
The SPring-8 experiments were carried out with the approval of the
Japan Synchrotron Radiation Research Institute (JASRI) (Proposal No.
2010B0084, 2011A0084).

\end{document}